\makeatletter
\newcommand{\dontusepackage}[2][]{%
  \@namedef{ver@#2.sty}{9999/12/31}%
  \@namedef{opt@#2.sty}{#1}}
\makeatother
\dontusepackage{subfigure}

\documentclass[]{article}

\usepackage{lmodern}
\usepackage{amssymb,amsmath}
\usepackage{ifxetex,ifluatex}
\usepackage[usenames,dvipsnames]{color}
\usepackage{fixltx2e} 
\ifnum 0\ifxetex 1\fi\ifluatex 1\fi=0 
  \usepackage[T1]{fontenc}
  \usepackage[utf8]{inputenc}
\else 
  \ifxetex
    \usepackage{mathspec}
    \usepackage{xltxtra,xunicode}
  \else
    \usepackage{fontspec}
  \fi
  \defaultfontfeatures{Mapping=tex-text,Scale=MatchLowercase}
  
\fi
\IfFileExists{upquote.sty}{\usepackage{upquote}}{}
\IfFileExists{microtype.sty}{%
\usepackage{microtype}
\UseMicrotypeSet[protrusion]{basicmath} 
}{}
\usepackage[margin=1.0in,bottom=1.5in]{geometry}
\usepackage[]{natbib}
\usepackage{listings}
\usepackage{graphicx}
\makeatletter
\def\maxwidth{\ifdim\Gin@nat@width>\linewidth\linewidth\else\Gin@nat@width\fi}
\def\maxheight{\ifdim\Gin@nat@height>\textheight\textheight\else\Gin@nat@height\fi}
\makeatother
\setkeys{Gin}{width=\maxwidth,height=\maxheight,keepaspectratio}
\usepackage{caption}
\usepackage{float}
\ifxetex
  \usepackage[setpagesize=false, 
              unicode=false, 
              xetex]{hyperref}
\else
  \usepackage[unicode=true]{hyperref}
\fi
\hypersetup{breaklinks=true,
            bookmarks=true,
            pdfauthor={},
            pdftitle={Time-domain Wavefield Reconstruction Inversion in a TTI medium.},
            colorlinks=true,
            citecolor=black,
            urlcolor=blue,
            linkcolor=black,
            pdfborder={0 0 0}}
\usepackage{mathtools}
\newcommand{\bm}{\mathbf{m}}

\newcommand{\bu}{\mathbf{u}}

\newcommand{\bd}{\mathbf{d}}
\newcommand{\bq}{\mathbf{q}}

\newcommand{\by}{\mathbf{y}}
\newcommand{\br}{\mathbf{r}(\bm)}

\newcommand{\F}{F(\bm)}
\newcommand{\A}{A(\bm)}
\newcommand{\R}{R}

\renewcommand{\L}{\mathcal{L}}

\newcommand{\LWRId}{\L_{\mathrm{TWRI}}}

\newcommand{\norm}[1]{\left\lVert#1\right\rVert}
\renewcommand{\ast}{^*}

\newcommand{\<}{\langle}
\renewcommand{\>}{\rangle}

\title{Time-domain Wavefield Reconstruction Inversion in a TTI medium.}
\author{Mathias Louboutin, Gabrio Rizzuti, and Felix J. Herrmann\\School of
Computational Science and Engineering, Georgia Institute of Technology}
\date{}

\begin{document}
\maketitle

\section{Abstract}\label{abstract}

We introduce a generalization of time-domain wavefield reconstruction
inversion to anisotropic acoustic modeling. Wavefield reconstruction
inversion has been extensively researched in recent years for its
ability to mitigate cycle skipping. The original method was formulated
in the frequency domain with acoustic isotropic physics. However,
frequency-domain modeling requires sophisticated iterative solvers that
are difficult to scale to industrial-size problems and more realistic
physical assumptions, such as tilted transverse isotropy, object of this
study. The work presented here is based on a recently proposed dual
formulation of wavefield reconstruction inversion, which allows
time-domain propagator that are suitable to both large scales and more
accurate physics.

\section{Introduction}\label{introduction}

Wave-equation based seismic imaging has become increasingly popular due
to its ability to produce detailed and accurate subsurface models. In
recent years, however, the limitations of Full Waveform Inversion (FWI)
have been widely acknowledged due to the cycle skipping issue that
arises with bandlimited data and lack of long offsets (thus low
frequencies). Simple geological settings, such as shallow water
sedimentary areas, have showcased the benefits of FWI, but more
challenging problems involve complex subsurface structures such as salt
bodies with strong anisotropy, which requires extensive manual
intervention for a consistently successful application. To address these
limitations, extended formulation have driven some of the most recent
research in seismic imaging. These methods rely on extra variables,
usually a wavefield \citep{vanLeeuwen2013GJImlm, vanleeuwenLocMin} as in
Wavefield Reconstruction Inversion (WRI) or time-space dependent source
\citep{huang2018volume, wang2016full} in the extended source method.
Extra unknowns are designed to absorb inaccuracies in the initial
background model by relaxing the physics, while additional constraints
ensure that this relaxation is ultimately driven to a physically
consistent scenario. One of the main challenges associated with extended
formulations is the necessity to solve for an augmented least-squares
system that, in the frequency domain, is only feasible for small to
mid-size problems with simple representations of the physics or for
industry-sized problems but of limited type of acquisition geometries
\citep{peters2019SEGans}. Conventional time-domain propagators, on the
contrary, do not share the same limitations.

This work is based on the time-domain formulation of WRI
\citep{rizzuti2019SEGadf} which relies on conventional time-stepping
\citep{devito-api, symes2015iwave}. Therefore, it straightforwardly
allows the adoption of a more accurate representation of the physics. In
this paper, we focus on the transverse tilted isotropic (TTI)
wave-equation \citep{zhang-tti, bubesatti2016, louboutin2018segeow}. The
inclusion of anisotropic effects are indeed crucial for the inversion of
field data. Note that frequency-domain methods do not enable TTI
anisotropy in full generality, with the exception of the limited
vertical transverse anisotropy (VTI) \citep{vtiwri}. Regardless, VTI WRI
still relies on iterative solvers for the augmented system, and is
therefore unfeasible for realistic sizes.

In this paper, we highlight the benefits of the time-domain WRI method
\citep{petersWRI, vanLeeuwen2013GJImlm, rizzuti2019SEGadf} for TTI
inversion and we demonstrate that TWRI can compensate for inaccuracies
in the anisotropic parameters in addition to compensating for a poor
initial model. We briefly introduce the dual formulation of WRI, then
present two examples that demonstrate that TTI WRI is feasible and more
robust than FWI with respect to modeling faulty assumptions.

\section{Methodology}\label{methodology}

In previous work \citep{rizzuti2019SEGadf}, we introduced the dual
formulation of WRI, which only requires conventional forward and adjoint
propagators in time domain and do not need the solution of the extended
wave equation. Time-domain WRI starts from the constrained least-squares
problem:
\begin{equation}
\min_{\bm,\bu}\dfrac{1}{2}\norm{\bq-\A\bu}^2\quad\mathrm{s.t.}\quad\norm{\bd-\R\bu}\leq\varepsilon,
\label{eq:denoise}
\end{equation}
 where $\bm$ represents the physical properties of interest and $\bu$ is
the associated wavefield. The data is denoted by $\bd$ and $\bq$ is the
(known) source term. The wave equation is denominated by $\A$ and $\R$
is receiver-restriction operator. We then derive the reduced Lagrangian
associated with Eq.~\ref{eq:denoise} (c.f
\citep{rockafellar-1970a, rizzuti2020EAGEtwri}) to obtain the TWRI
objective function:
\begin{equation}
    \begin{split}
    & \max_{\by}\min_{\bm}\LWRId(\bm,\by),\\
    & \LWRId(\bm,\by)=-\dfrac{1}{2}\norm{\F^{\ast}\by}^2+\<\by,\br\>-\varepsilon\norm{\by},
    \end{split}
\label{lagr_red}
\end{equation}

where the forward operator $\F=\R\A^{-1}$ is the forward modeling
operator, and $\br=\bd - \F \bq$ the data residual for the model $\bm$.
This dual problem has two unknowns, parameters $\bm$ (squared slowness),
and variables $\by$(having the same size of data). One of the advantages
of the formulation in Eq.~\ref{lagr_red} is that, unlike conventional
WRI, the additional variable is of a manageable size (compared to
wavefields). To update the model $\bm$, we calculate the derivative:
\begin{equation}
\nabla_{\bm}\LWRId=-J[\bm,\bq+\F^{\ast}\by]^{\ast}\by,
\label{lagr_red_gradm}
\end{equation}
 where $J$ is the conventional FWI Jacobian operator for an extended
source $\bq+\F^{\ast}\by$. It is straightforward to extend the previous
acoustic implementation to the Transverse Tilted Isotropic case
\citep[TTI,][]{zhang-tti, duveneck, louboutin2018segeow}, just by simply
replacing $\F$ with its TTI counterpart.

Previous work have demonstrated that TWRI behaves more robustly than FWI
with respect to local minimum issues in the purely acoustic case. Here
we illustrate that the edge of WRI over FWI still holds for TTI and,
more importantly, when the assumed physics do not match with the
collected data.

\section{Numerical experiments}\label{numerical-experiments}

These two examples aim to elucidate two main advantages of TWRI. First,
since we use a time-domain formulation that only necessitates the
implementation of standard forward and adjoint propagators, we are able
to implement TTI TWRI by simply replacing the acoustic time-stepper with
an anisotropic version thereof. Moreover, thanks to
\href{https://github.com/devitocodes/devito}{Devito}
\citep{devito-api, devito-compiler} and
\href{https://github.com/slimgroup/JUDI.jl}{JUDI}\citep{witteJUDI2019},
these propagators are implemented in a simple high-level way and benefit
from its highly efficient just-in-time compiler. Finally, we show that
TTI TWRI mitigates cycle skipping either with true anisotropy parameters
and a kinematically inaccurate initial model, or with inaccurate
anisotropic parameters and a fair initial model. Finally, we verify a
known secondary advantage of WRI that is its robustness to inaccurate
water layer velocity and ocean bottom position.

We concentrate on three increasingly difficult settings over these
examples:

\begin{enumerate}
\def\labelenumi{\arabic{enumi}.}
\itemsep1pt\parskip0pt\parsep0pt
\item
  Invert the TTI data with a TTI wave-equation and assuming the true
  anisotropy parameters are known.
\item
  Invert the TTI data with a TTI wave-equation with error introduced in
  the anisotropic parameters.
\item
  Invert the TTI data with an acoustic wave-equation.
\end{enumerate}

The first two cases demonstrate the TWRI mitigates the sensitivity to
cycle skipping associated with both the velocity and anisotropy errors.
The third case demonstrates that TWRI can compensate for a numerical
representation of the physics that does not correspond to the physics of
the observed data.

\subsection{Gaussian lens}\label{gaussian-lens}

This first example follows \citep{huang2018volume} with a 2D singularity
model with a constant initial model. This model in known to be cycle
skipped and demonstrate the capability of TWRI to obtain a better update
direction than conventional FWI. To further highlight the robustness
with respect to the Thomsen parameters\citep{thomsen1986}, we fix these
to zero and compute the gradient. This parameterization shows that TWRI
can compensate for anisotropy errors for the computation of the gradient
with respect to the squared slowness.

\begin{figure}
\centering
\includegraphics[width=1.000\hsize]{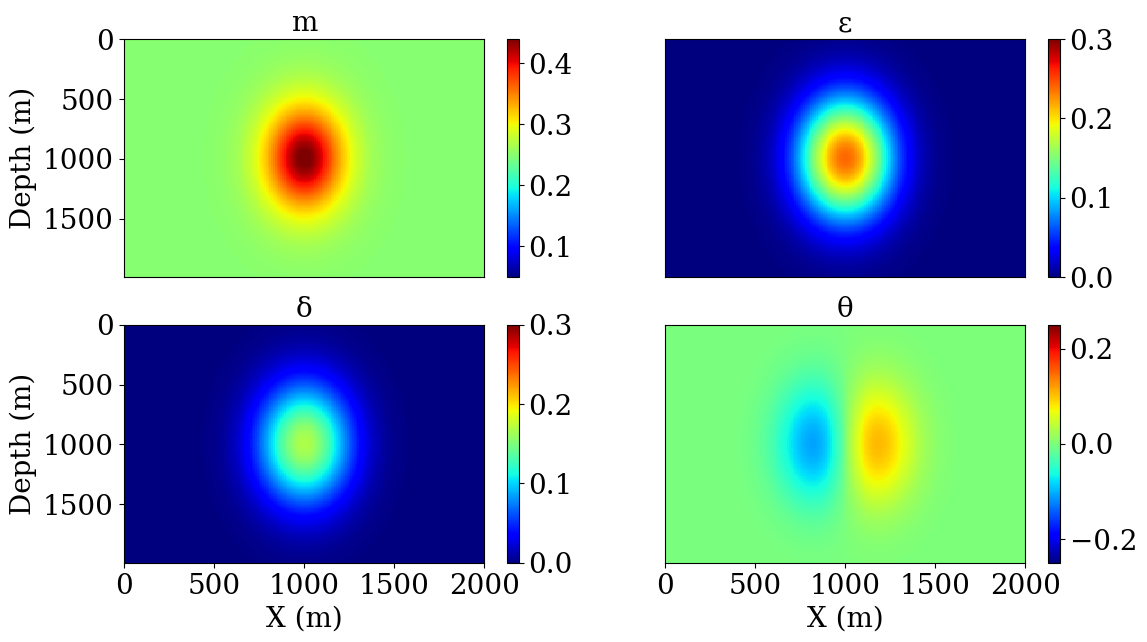}
\caption{Squared slownes, Thomsen parameters and dip angle for the BG
Compass model.}\label{GLmodel}
\end{figure}

\begin{figure}
\centering
\includegraphics[width=1.000\hsize]{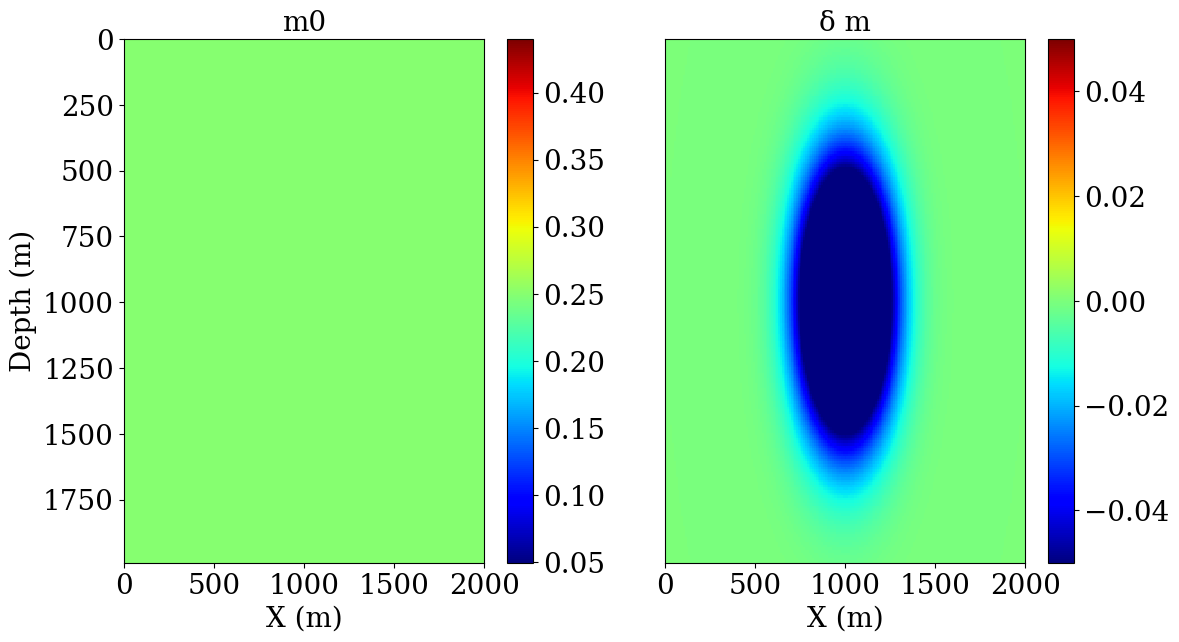}
\caption{Initial models and true perturbation for the Gaussian lens
model. The initial model is a constant velocity and the true
perturbation is the velocity singularity.}\label{GLdm}
\end{figure}

We show the initial model (constant velocity) and true perturbation in
Figure~\ref{GLdm}, and we expect to see the gradient update to have the
same sign as the true perturbation in Figure~\ref{GLdm}.

\begin{figure}
\centering
\includegraphics[width=1.000\hsize]{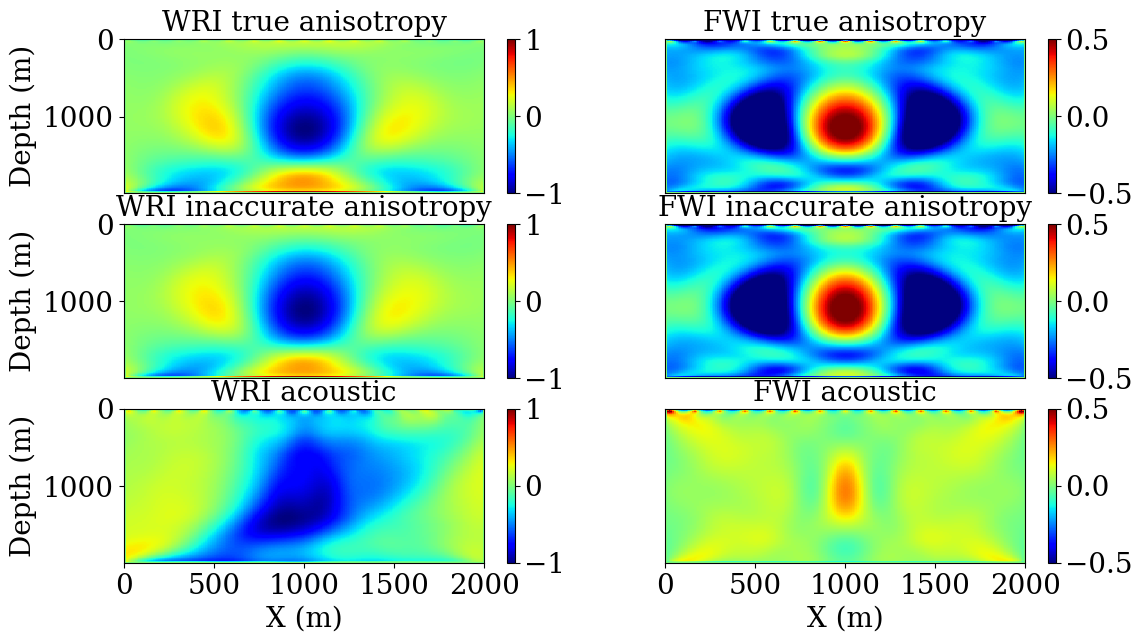}
\caption{Gradients obtained with the anisotropic data for the three
different inversion settings.}\label{GLgrad}
\end{figure}

The gradient obtained with each of the three cases are displayed on
Figure~\ref{GLgrad}. We first see that, as expected from previous work,
FWI does not produce a good update direction as the sign is flipped
compared to the true perturbation (FIgure~\ref{GLdm}) and optimizations
algorithms will not converge to a good solution. On the other hand, we
can see that in all three cases, the update direction obtained with TWRI
is consistent with the true perturbation and will lead to a good model
reconstruction. One interesting observation is that TWRI correctly
handles the errors in the anisotropy, including for the complete absence
of anisotropy in the modeling kernel. Such result is encouraging in
light of more realistic examples.

\subsection{BG Compass}\label{bg-compass}

\begin{figure}
\centering
\includegraphics[width=1.000\hsize]{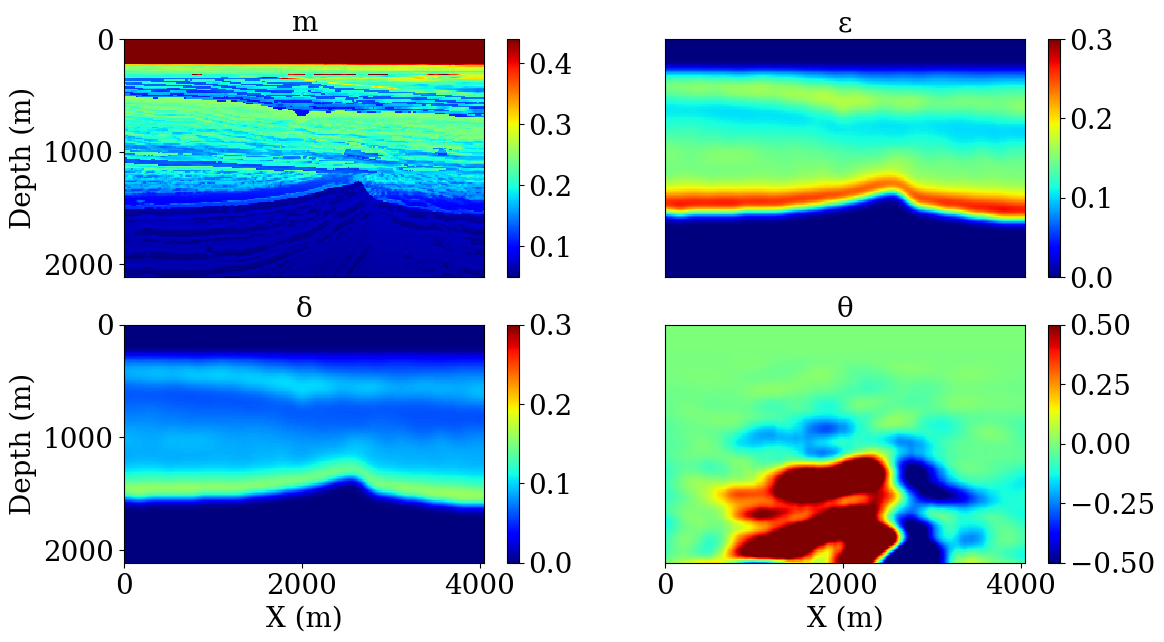}
\caption{Squared slowness, Thomsen parameters and dip angle for the BG
Compass model.}\label{BGback}
\end{figure}

The second example involves the BG Compass model. The Thomsen parameters
are synthesized from the velocity model, while dip angles are inferred
from the orientation of the layers. We show the TTI parameters in
Figure~\ref{BGback}. This model is notoriously difficult due to the
velocity kick back (situated at around the one kilometer depth on
Figure~\ref{BGback}) that prevents turning waves from traveling back to
the surface.

\begin{figure}
\centering
\includegraphics[width=1.000\hsize]{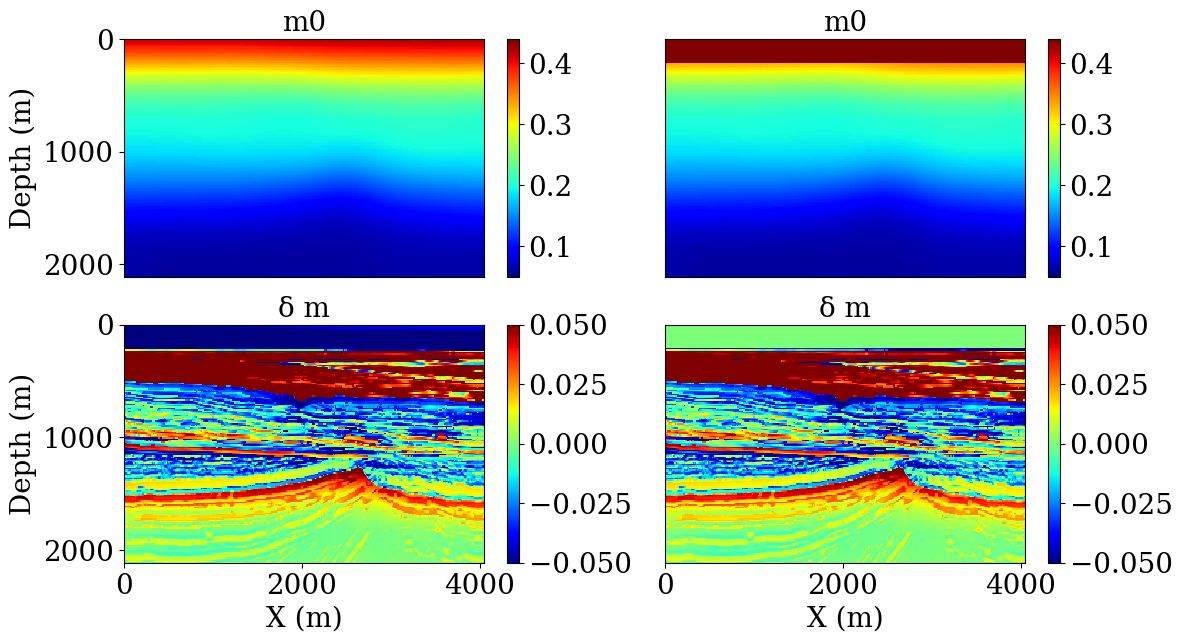}
\caption{Initial models and true perturbation for the two test cases.
The two initial models are smoothed version of the true model with a 20
points gaussian filter, and the right initial model has the true water
layer speed and ocean bottom position.}\label{BGdm}
\end{figure}

This experiment is divided in two settings. On one hand, we assume the
water velocity and depth of the water layer to be known and look at the
first update computed with TWRI and FWI. In the second case, we do not
assume any knowledge of the water layer, that is known to be delignated
by WRI in the acoustic case, and once again look at the first FWI and
TWWRI update. These two initial models and the true perturbation
associated with it are on Figure~\ref{BGdm}, and a good update is
expected to correlate with the true perturbation well while a cycle
skipped update would have the wrong sign in most areas in particular for
the velocity kick-back part.

\begin{figure}
\centering
\includegraphics[width=1.000\hsize]{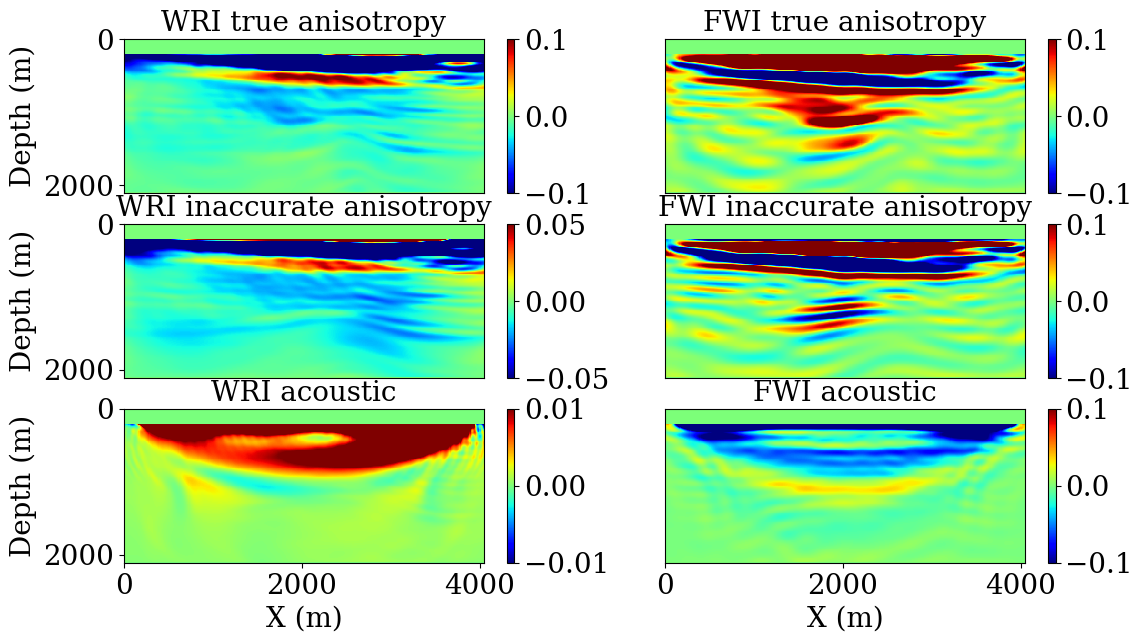}
\caption{Gradients obtained with the anisotropic data and an initial
model without the correct water layer.}\label{BGgrad}
\end{figure}

The results obtained with the correct water velocity are shown in
Figure~\ref{BGgrad}. Similarly to the previous example, TWRI succeeds to
capture most of the features of the true perturbation while FWI displays
major differences. The most difficult part of the model to image is the
middle section around 1 km depth. The velocity kick-back makes the
inversion very difficult as the turning wave are diving deeper into the
model rather than going back up towards the receivers. We can see that
TWRI perfectly captures this velocity kick back while FWI only succeeds
to match partially the shallower part of the model.

In the second part of this experiment, the results obtained with the
incorrect water velocity are shown on Figure~\ref{BGgrad}. As expected
from previous results \citep{petersWRI}, TWRI provides a correct update
direction in the water layer while still featuring the attributes
necessary for inversion such as the previously mentioned velocity
kick-back.

\begin{figure}
\centering
\includegraphics[width=1.000\hsize]{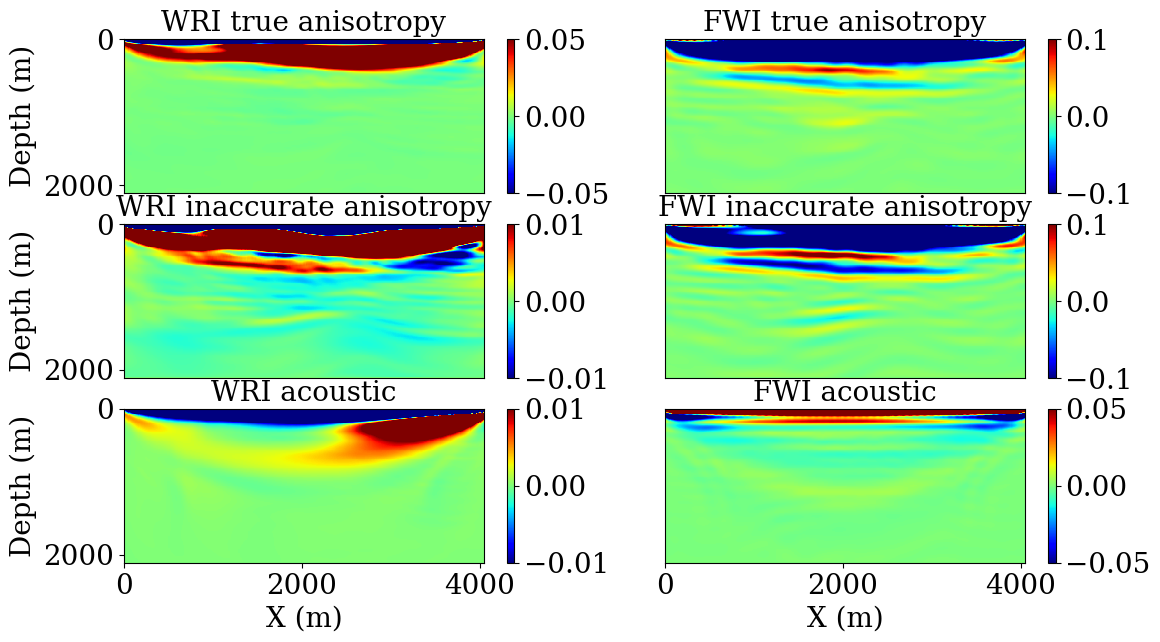}
\caption{Gradients obtained with the anisotropic data and an initial
model with the correct water layer.}\label{BGgradw}
\end{figure}

These examples and related software can be found at
\href{https://github.com/slimgroup/Software.SEG2020}{TTIWRI} in our
reproducibility repository
\texttt{https://github.com/slimgroup/Software.SEG2020}.

\section{Discussion \& conclusions}\label{discussion-conclusions}

In this work, we presented an application of TWRI to a realistic
inversion scenario, by including TTI physics. Because our work is based
on time-domain modeling, we can leverage on state of the art anisotropic
propagators. The extension to anisotropy is not trivial in the frequency
domain due to the current limitations of iterative solvers for
large-scale, coupled PDEs. We experimented that TWRI not only produces
more qualitative results that conventional FWI when data and modeling
physics matches, but also fairs better with respect to modeling
inaccuracies, e.g.~when the inverted data presents anisotropic effects
but an acoustic medium is postulated. While these preliminary results
are encouraging, future work will focus on a more thorough validation of
TWRI for TTI media and applications to 3D examples.

\section{References}\label{references}

\bibliography{tti_wri}

\end{document}